\documentclass[twocolumn,prl,aps,showpacs]{revtex4}
\usepackage{graphicx}

\mathchardef\ogon="012C%
\newcommand{\as}{a\kern-0.22em\lower.40ex\hbox{$_{\ogon}$}}

\begin{document}
%\wideabs{
\title{Atomic Bose and Anderson glasses in optical lattices}
\author{B. Damski$^{1,2}$, J. Zakrzewski$^{1}$, L. Santos$^{2}$,
P. Zoller$^{2,3}$, and M. Lewenstein$^{2}$}
\address{
(1)Instytut Fizyki  im. M. Smoluchowskiego, Uniwersytet
Jagiello\'nski,
Reymonta 4, PL-30 059 Krak\'ow, Poland \\
(2)Institut f\"ur Theoretische Physik, Universit\"at Hannover, D-30167
Hannover, Germany\\
(3)Institut f\"ur Theoretische Physik, Universit\"at Innsbruck, A-6020,
Innsbruck, Austria}

\begin{abstract}
An ultra cold atomic Bose gas in an optical lattice is shown
to provide an ideal system for the controlled analysis of {\em disordered} Bose lattice gases.
This goal may be easily achieved under the current experimental conditions,
by introducing a pseudo-random potential created
by a second additional lattice or, alternatively,
by placing a speckle pattern on the main lattice.
We show that for a non commensurable filling factor, in the strong interaction limit,
a controlled growing of the disorder drives a dynamical transition from superfluid
to Bose-glass phase. Similarly, in the weak interaction limit,
a dynamical transition from superfluid to Anderson-glass phase may be observed.
In both regimes, we show that even very low-intensity disorder-inducing lasers cause
large modifications of the superfluid fraction of the system.

\end{abstract}
\maketitle

%}
%shortcuts:
%Anderson localized = Anderson glass == AG, Bose glass == BG
%SF, MI

At zero temperature, the bosonic lattice gases may undergo a quantum phase
transition~\cite{sachdeev} from
a superfluid (SF)  to an insulator ~\cite{fisher}.
In the absence of disorder there exist Mott insulator
(MI) states, characterized by a fixed (integer) number of bosons per lattice site,
a gap in the excitation spectrum, and vanishing superfluid fraction and compressibility.  In the presence of disorder, an additional insulating phase, so-called
Bose-glass phase (BG), may occur \cite{fisher}. This phase presents no
superfluid fraction, a gapless excitation spectrum, and a finite compressibility.
The SF-BG transition has been recently experimentally studied in various physical
systems \cite{experiment}, and attracts a continuous theoretical interest
\cite{batrouni,krauth,kiesker,lee}.
In particular, the possibility of a direct MI-SF transition in the presence of
disorder remains a controversial issue.

The nature of the disorder-induced insulator phases depends
on the interplay between hopping, nonlinear interactions and
disorder. In the strong-interaction regime, the cooperation of interactions
and disorder
leads to the appearance of a BG phase. For weak interactions, the disorder leads to
an Anderson-type insulator, or Anderson-glass (AG)~\cite{batrouni}. 
In the latter case, contrary to the BG phase, the interactions tend to delocalize and
therefore compete with the disorder.

The detailed analysis of these properties demands an
experimentally accessible system in which the disordered Bose
lattice gases could be studied in a controlled way. One of the
aims of this paper is to show that this goal can be relatively
easily accomplished by using cold Bose gases in optical lattices,
for which the development of cooling and trapping techniques
allows a large degree of control. Recently, Greiner {\it et al.}
\cite{greiner}, following the theoretical suggestion of Jaksch
{\it et al.} \cite{jaksch}, have observed the SF to MI quantum
phase transition in an optical (non disordered) lattice loaded
with $^{87}$Rb atoms.

In the case of an optical lattice a pseudo-random potential can be
dynamically generated by growing on an already  existing (main)
optical lattice a second (additional) one with a different
wavelength \cite{jaksch,burnett,diener}.  Based on this idea, we
study in this Letter the dynamical generation  of the BG and
AG phases in a many-body system under realistic conditions. In
addition, we discuss the alternative generation of a truly random
potential by superimposing a speckle pattern. In the first part of
this Letter, we consider the regime of strong interactions where
the filling factor is non-integer, showing
that the adiabatic turn-on of the
disorder may lead to a dynamical transition into the BG phase. In the last part of the Letter,
we analyze the weak interaction regime, and in particular the possibility to achieve
a dynamical transition into the AG phase. For both regimes of interactions,
the SF fraction~\cite{lieb} is calculated.
We show that under realistic conditions even very low-intensity
disorder-inducing lasers lead to a dramatic reduction of the SF fraction, indicating
the appearance of BG, or AG phases.

We consider an ultracold Bose gas in a 2D optical lattice.
We analyze the case of $^{23}$Na, but our results also apply to other species.
We assume the atoms as tightly confined in the transversal ($z$) direction by a harmonic trap
of frequency $\omega_z/2\pi= 6$kHz, so that the wavefunction in $z$ remains the
Gaussian ground state.
No additional harmonic confinement is assumed in the $xy$ plane.

The optical lattice is formed by the main laser beams, whereas additional
lasers are responsible for the introduction of (quenched)
pseudo-disorder. The optical potential has thus the following form:
\begin{equation}
\label{lattice}
V({\bf r})= V_l({\bf r}) + V_r({\bf r}),
\end{equation}
where ${\bf r}=[x,
y]$, the main lattice is $V_l({\bf r})= V_0(\cos^{2}(kx)+\cos^{2}(ky))$, and
the secondary one is $V_r({\bf r})= V_1(\cos^{2}({\bf k_1 r})+ \cos^{2}({\bf {k_2} r}))$. The
main (additional) beams intensity and/or detuning controls the value of
$V_0$ ($V_1$). In the following we assume
$k_1=k_2 \neq k$,  and the directions
${\bf {k_1}}\propto [-0.5, 1]$ and ${\bf {k_2}}\propto [-1, 0.5]$.
The pseudo-randomness of the
potential is determined by $q=k_1/k$,
i.e. the ratio between the wavelengths  of the main ($\lambda=2\pi/k$)  and the
additional lattices. For commonly used NdYag and TiSapphire lasers
$q=1064/795=1.338$. 
We note that a pseudo random lattice can also be achieved by
splitting off part of the main laser beams and creating the
additional incommensurate lattice by interfering these light beams
at an angle.

We assume that the energies involved in the system are much smaller than the
energy separation between the first and the second band of the lattice,
and consequently
we can reduce our analysis to the first band. In that case,
the physics of the atomic lattice Bose gas is governed by the Bose-Hubbard Hamiltonian~\cite{jaksch}:
\begin{equation}
\label{H1}
H=-\sum_{<i,j>}J_{ij}a^{\dagger}_ia_j +
\frac{U_0}{2}\sum_{i}n_i(n_i-1)+  \sum_{i}W_i n_i.
\end{equation}
$J_{ij}=J_0+\delta J_{ij}$ are the tunneling (hopping) coefficients between nearest neighbors. 
They slightly differ from $J_0$ \cite{jaksch}
by a correction of the form:
$ \delta J_{ij}=-\int d^3r  w^{\star}({\bf r-r_i})V_r({\bf r})w({\bf r-r_j})$, where
$w({\bf r-r_j})$ are the Wannier functions for the lowest energy band.
Clearly, the contributions of the additional beams to the tunneling, vanish
on ``average'',  $<\delta J_{ij}>\approx 0$.
In the following we consider $V_0=25 E_{R}$ and $V_1\simeq0.05E_R$,
where $E_{R}$ is the photon recoil energy.
In this case we have checked numerically that $|\delta J_{ij}/J_0| < 0.1\%$, hence
the major contribution to $J_{ij}$ comes from $J_0$, and the model
reduces to the ordinary Bose-Hubbard one with constant tunneling \cite{batrouni,krauth,lee,kiesker}.
In the Hamiltonian~(\ref{H1}),
$U_0\propto a$  is the coupling constant~\cite{jaksch} for the
interparticle interactions
($a$ is the scattering length), and
\begin{equation}
W_i= \int d^3r w^{\star}({\bf r-r_i})V_r({\bf r})w({\bf r-r_i}),
\label{vi}
\end{equation}
are the pseudo-random on-site energies, which, as discussed below, may
introduce significant effects even for very small $V_1/V_0$.

The disorder-induced phases are characterized by a
vanishing SF fraction, which is determined studying
the system sensitivity to changes of boundary
conditions. To this aim, we employ the {\it boost} method \cite{lieb},
resulting in substitution in ~(\ref{H1}): $J_{ij}\rightarrow J_{ij} e^{i\varphi_{ij}}$. 
The angles $\varphi_{ij}$ are defined as
follows: if $i=(x_i, y_i)$ and $j=(x_j, y_j)$, then for
$y_i=y_j$, $\varphi_{ij}= {\rm sign}(x_i-x_j)\varphi/M$,  and else $\varphi_{ij}=0$,
where $M$ is the lattice size in the $x$-direction.
Physically, this choice of $\varphi_{ij}$ corresponds to a constant current per lattice,
proportional to $\varphi$, in the positive $x$ direction.
The SF fraction is then obtained from the corresponding ground-state energy
$E(\varphi)$ as
\begin{equation}
\label{syf}
 \rho_s= \frac{M^2}{N}\frac{E(\varphi)-E(0)}{J_0 \varphi^2},
\label{rhos}
\end{equation}
where $N$ is the number (mean) of atoms.
Similar expression can be easily derived for a 1D case.
In the following,
we denote by $\Upsilon$ the ratio of $N$ to the number of lattice sites.
The ground state %energy
is obtained from the minimization of
$\langle \Psi_{MF}|H-\mu N|\Psi_{MF}\rangle$, where $\mu$ is the
chemical potential. We employ the Gutzwiller ansatz
$|\Psi_{MF}\rangle=\prod_{i}\sum_n^{\infty}f_n^{(i)}|n\rangle_{i}$, where
$f_n^{(i)}$ are the amplitudes of having $n$ atoms at a $i$-th lattice site.
We have observed, that it is numerically easier
to calculate the ground state
using static methods for a non-disordered and non-twisted case ($V_1=0$,
$\varphi=0$),
and then adiabatically evolve such a state, first  to $\varphi\neq0$, and then
for a constant $\varphi$ to $V_1\neq0$ (first evolving $V_1$ and then
$\varphi$ should give the same result).
The evolution is performed by means of the dynamical
Gutzwiller approach ~\cite{dynam}, in which we solve the equations
\begin{eqnarray}
\label{T}
i\dot{f}^{(i)}_n &=& \left[\frac{U_{0}}{2}n(n-1)+n W_i\right]f^{(i)}_n+
\nonumber \\
&&\Phi^{\star}_i \sqrt{n+1} f^{(i)}_{n+1}+ \Phi_{i} \sqrt{n} f^{(i)}_{n-1},
\label{dyn}
\end{eqnarray}
where
$\Phi_{i}=-\sum_{<i,j>} J_{ij}e^{i\varphi_{ij}}
\langle\Psi_{MF}|a_j|\Psi_{MF}\rangle$.

Another useful quantity characterizing the state of a Bose lattice gas
is the condensate fraction,  defined as the highest
eigenvalue of the one particle density matrix, $\rho_{ij}=\langle
\Psi_{MF}|a_i^{\dagger}a_j|\Psi_{MF}\rangle$, divided by the
number of particles~\cite{pitaevskii}. This quantity is
important for experiments, since it determines the phase coherence, and thus
the contrast in interference measurements \cite{greiner}.

\begin{figure}[hbp]
\begin{center}
 \includegraphics[angle=-0,scale=0.24, clip=true]{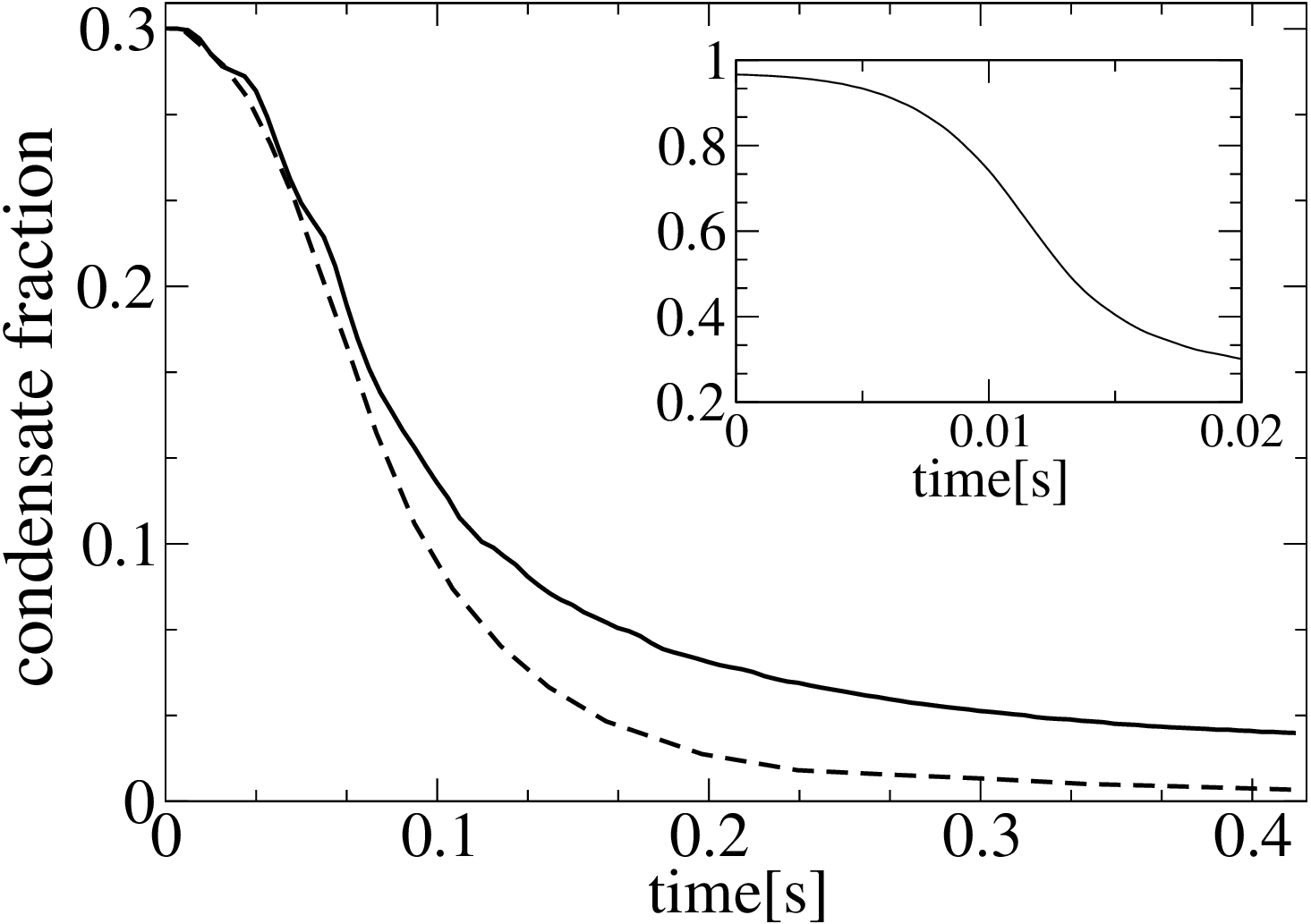}
\caption{Condensate fraction (solid line) during the dynamical SF to BG transition
induced by
a superimposed additional optical lattice, for
$^{23}{\rm Na}$ atoms placed on the main $40\times 40$ lattice with $\lambda=1064$nm, $q=1.338$,
$\Upsilon=0.75$, $V_0/E_R=25$, and $V_1(t)/E_R= 0.059t/T$ where $T=0.42s$ is the
total time of evolution. The inset shows the first preparation step (see text),
for which $V_0(t)= V_{SF}+(V_{MI}-V_{SF})t/T$ ($V_1(t)=0$),
where $V_{SF}/E_R=7$, $V_{MI}/E_R=25$, and $T=0.02$s.
The dashed line presents the SF fraction.
}
\label{bgtran}
\end{center}
\end{figure}

In order to study the dynamical transition into the BG phase,
it is convenient to prepare the system in the SF phase in the presence
of dominating interactions and weak tunneling, when $\Upsilon$ is non-integer.
Therefore, we consider an initial system deeply in the SF regime, with almost
$100\%$ of SF fraction and $\Upsilon=0.75$.
Then, the intensity of the main laser is adiabatically increased
in $20$ms,
obtaining a very large $U_0/J_0 \approx 70$.
Both, the condensate and the SF  fraction decrease during this
process down to approximately $30\%$, being non-zero only due to non-integer
value of $\Upsilon$ (see inset of Fig.~\ref{bgtran}).
As a next step, the disorder is turned-on adiabatically
in about $0.5$s, by switching on the additional laser beams.
The condensate and also the SF
 fraction decrease dynamically during this process (Fig.~\ref{bgtran}).
Ultimately, the condensate fraction does not tend to $0$, but to a very small
value of about $2\%$, due to the finite size of the systems and the
approximate character of
the Gutzwiller approach \cite{zwerger,remark}. In contrast, the SF
fraction tends to zero faster.
Note that the superfluidity is rapidly lost, although at any time the
additional lattice is very much weaker than the main one. This fact could
at first glance seem surprising, but it is due to the
small values of $J_0$ and $U_0$  ($10^{-3}E_R$ and $7\times 10^{-2}E_R$,
respectively). Thus even for
$V_1$ being of the order of few percent of $E_R$ ($V_1/V_0 \ll 1$),
the value of $|W_i|/U_0\sim 1$. Together with the low value of $J_0/U_0$,
this explains why the system enters the BG phase for such a weak
additional lattice.

We stress at this point that our calculations do not include an additional 
inhomogeneous trapping in the $xy$ plane, which, for the low $J_0$ 
we consider, may result in the formation of MI domains \cite{muramatsu} if
the filling factor at the trap center is larger than $1$. 
For sufficiently shallow traps, the central BG region should dominate 
the physics for a finite disorder. For filling factors 
lower than 1, the systems is expected to be fully in the BG phase.

\begin{figure}[hbp]
\begin{center}
 \includegraphics[angle=-0,scale=0.24, clip=true]{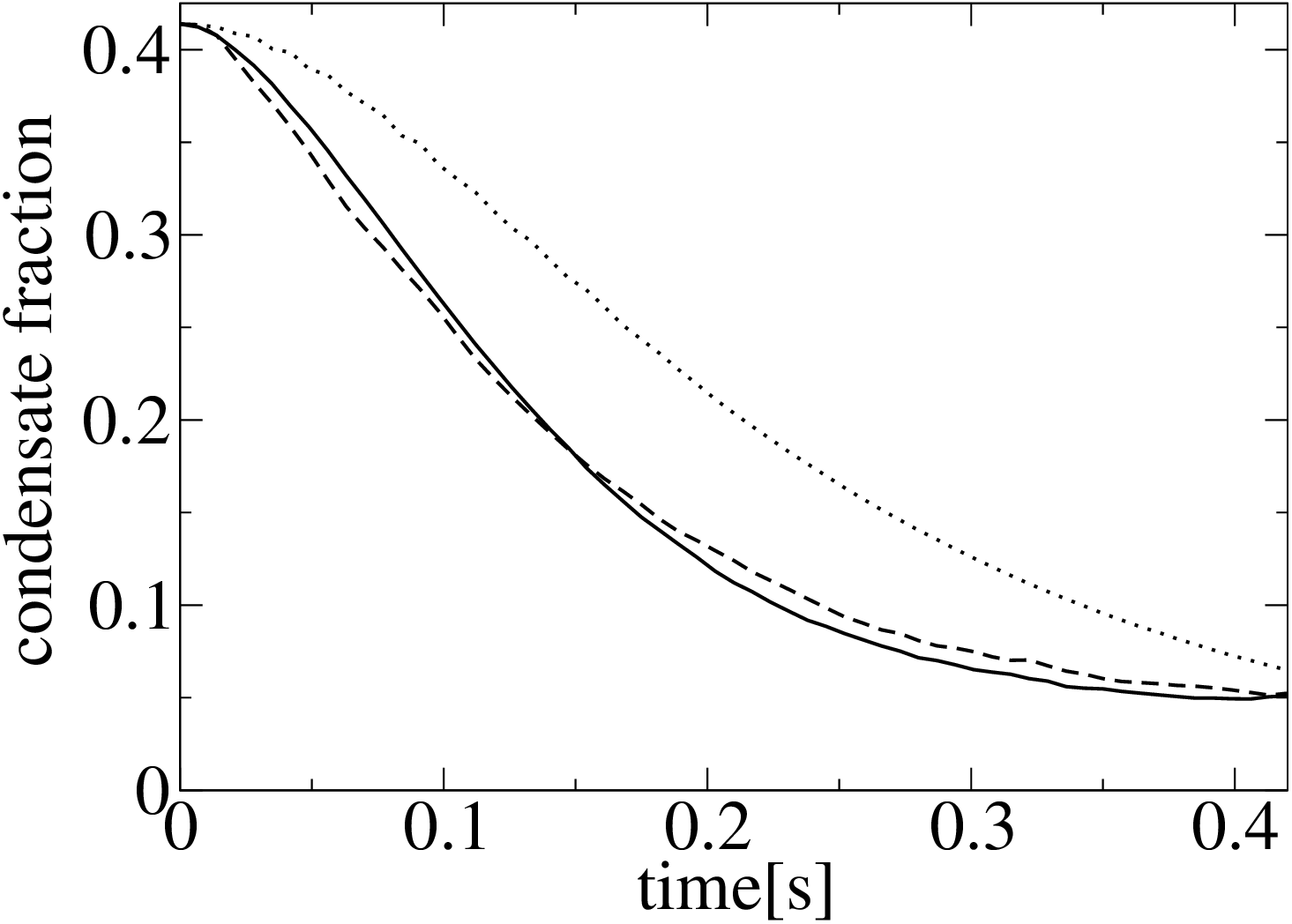}
\caption{Condensate fraction during the dynamical SF to BG
transition induced by
a superimposed speckle pattern, for
$^{23}{\rm Na}$ atoms placed on the main $40\times 40$
lattice with $\lambda=1064$nm,
$\Upsilon=1.5$, $V_0/E_R=25$, and $V_2/E_R=0.048t/T$, with $T=0.42$s.
The solid (dashed) line refers to speckles of average size $\Gamma=0.34 \lambda$
($\Gamma=1.37 \lambda$).
For comparison, the dotted line shows the same transition
in the quasi-disordered case (parameters as in Fig.~\ref{bgtran}, but
$\Upsilon=1.5$).
We expect that similarly as in Fig. ~\ref{bgtran} the decrease of the SF fraction is
faster than that of the condensate one.
}
\label{spec}
\end{center}
\end{figure}

It is interesting to compare results obtained with the quasi-disordered
perturbation induced  by the additional lattice, and those achieved using
a purely random optical potential coming from a speckle pattern.
We generated speckle pattern in a way described in \cite{speckle} and
checked that it gives correctly both the autocorrelation function and the
probability distribution.
The speckle pattern induces a potential
$V_s(\vec{r}\;)$, which is characterized by its mean value
$V_{2}=\langle V_s(\vec{r}\;)\rangle$, and by the average speckle size $\Gamma$.
As above,  the disorder-induced
 corrections to the tunneling are minor, and the most
significant contribution of the speckle potential appears in the form of $W_i$ coefficients
~(\ref{vi}), defined in this case by means of the $V_s(\vec{r}\;)$ potential.
Not surprisingly, values of $V_2$ comparable to those previously considered for $V_1$,
lead to a transition from the SF to Bose glass phase.
We have investigated this issue for $\Gamma= 0.34$, $1.37$, $2.75 \lambda$.
The obtained results are similar to the ones resulting from quasi-random
perturbation generated by additional beams (see Fig.~\ref{spec}).

\begin{figure}[hbp]
\begin{center}
\includegraphics[angle=-0,scale=0.32, clip=true]{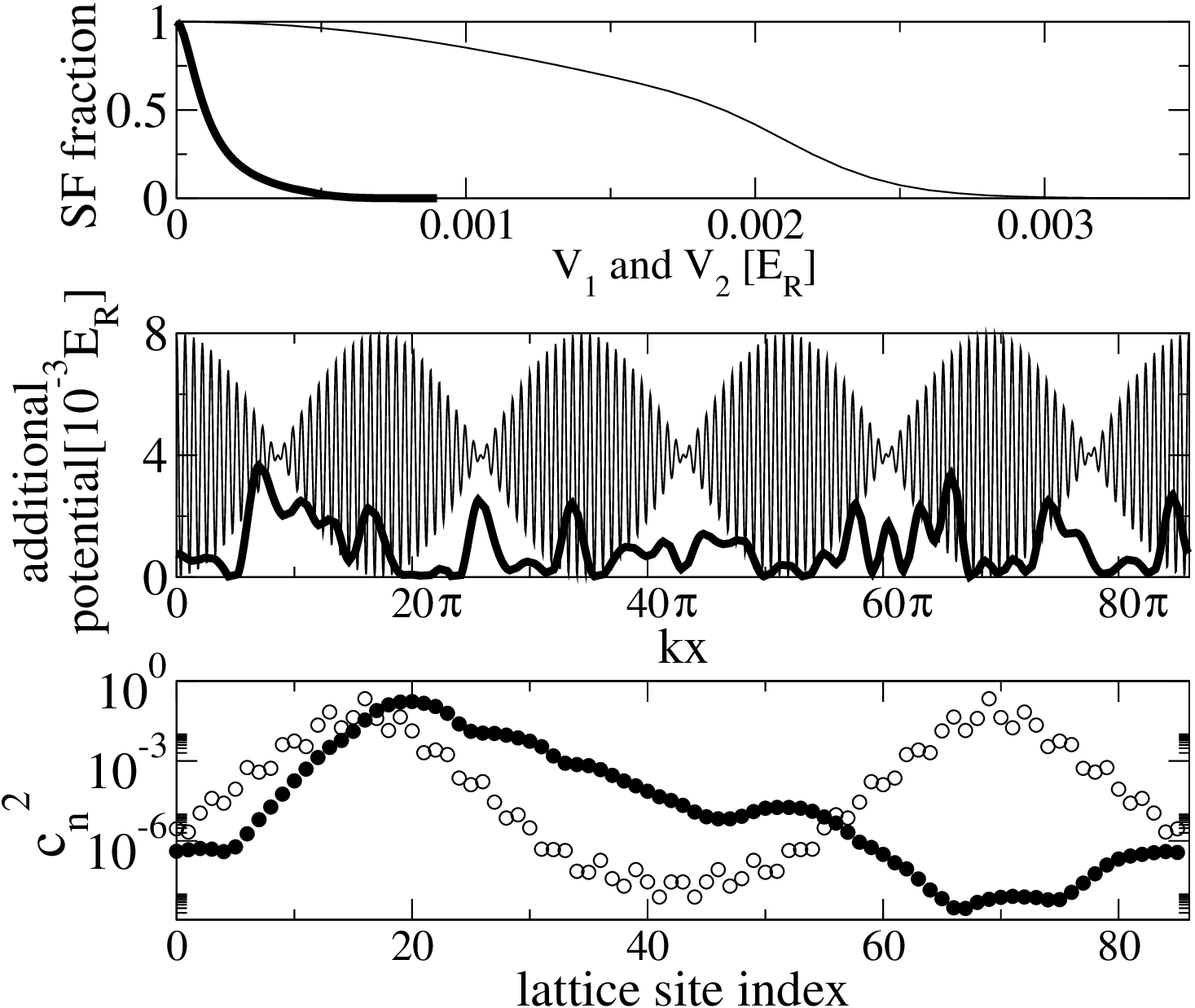}
\caption{
Transition from the  SF to AG phase for
$^{23}{\rm Na}$ atoms placed on a 1D main lattice of $86$ sites, with $\lambda=1064$nm and $V_0/E_R=16$.
Upper plot: superfluid fraction vs. amplitude of the disordered potential.
The thick curve indicates the speckle-induced transition ($\Gamma=1.34\lambda$), whereas the normal line refers to the transition generated by a superimposed
%incommensurate
lattice with $k_1/k=1.338$, $k_2/k=1.396$ (see text).
Middle plot: dependence of $V_r(x)$ (normal line) and
$V_s(x)$ (thick line) for $V_1/E_R=0.004$ and $V_2/E_R=0.001$.
Lower plot: occupation numbers of the ground state of the Hamiltonian~(\ref{H2}) in the presence of
speckles (filled circles) and additional beams (hollow circles), for
$V_1/E_R=0.004$ and $V_2/E_R=0.001$.}
\label{agtran}
\end{center}
\end{figure}

In the final part of this Letter, we discuss the non-interacting regime, where
the adiabatic turn-on of the disorder can lead to a dynamic transition
into the AG phase.
The weakly interacting regime, and even the non-interacting one,
could be achieved
by reducing the $s$-wave scattering length by means of Feshbach
resonances~\cite{cornell}.
For simplicity we consider a 1D optical lattice case where the perturbation
is either generated by two non-commensurate standing waves,
$V_r(x)=V_1(\cos^2(k_1x)+\cos^2(k_2x))$ with $k_1\neq k_2$,
or by a laser speckle potential $V_s(x)$
characterized by a mean value $V_2$.
The Hamiltonian (\ref{H1}) reduces then to the form
\begin{equation}
\label{H2}
H=-\sum_{<i,j>}J_{ij}a^{\dagger}_ia_j + \sum_{i}W_i n_i,
\end{equation}
which becomes the famous Anderson's hopping model ~\cite{haake}, if the
$W_i$ coefficients are {\em randomly} distributed.
Similarly as above $J_{ij}$ are almost unchanged by the presence of disorder.
The single-particle character of the non-interacting problem
greatly simplifies the
calculations and allows for an exact treatment.
In the absence of interactions, 100\% of atoms condense at zero temperature,
each in a single particle state
$|\Psi\rangle=\sum_n c_n a^{\dagger}_n|{\rm vac}\rangle$.
We have analyzed the
dynamical evolution of the amplitudes $c_n$ during the turn-on of the disorder
as well as their ground state distributions---see Fig.~\ref{agtran}.

Due to the finite size of the system, which implies  discreteness of the excitation spectrum, the application of the boost method in the absence of disorder
results in the 100\% SF fraction. When the
typical energy of disorder becomes comparable to the energy separation between the energy levels in the absence of disorder, the SF fraction is expected
to vanish (see %upper plot in
Fig.~\ref{agtran})
We clearly observe a transition from a delocalized SF state to an Anderson
localized one, where the occupation probabilities of neighboring sites decrease
{\it exponentially} with the distance from the localization 
centers~\cite{haake}. From the upper plot in Fig.~\ref{agtran} we conclude 
that in order to drive
the system into the AG
phase, smaller intensities are needed for the speckle pattern
than for the superimposed incommensurable lattice, since as expected
the speckle pattern results in a  ``more
disordered'' distribution of the $W_i$ coefficients
\cite{notka1}
(see middle plot in
Fig.~\ref{agtran}).
It is especially worth to stress that in the case of a superimposed incommensurable lattice
the periodicity of the lattice reflects itself in periodically  {\it Anderson-localized}
domains,
but the localization phenomena is still present, as shown in
Fig.~\ref{agtran}. We observed
the same quantitative phenomena also for ratios $k_1/k$ and $k_2/k$ different
than the ones listed in the caption of Fig.~\ref{agtran}.
We have performed also time-dependent simulations based on Eq.~(\ref{H2}) to
determine the time scale for the adiabatic transition into the AG phase.
For a main lattice
wavelength $\lambda=1064$nm the adiabatic evolution lasts few seconds.
This may be shortened using higher frequency lattice beams
since evolution time  scales as $\lambda^2$.

The experimental observation of the SF to BG (or to AG) transition
should be relatively easy to accomplish in a setup as that of Ref.~\cite{greiner}.
The BG and AG phases can be detected 
by observing the interference pattern after removing the lattice
for different intensities of the superimposed laser beams.
The insulator character of the BG and AG phases will be revealed 
by the disappearance of the interference fringes. 
The phases can be additionally characterized by measuring their gapless  
excitation spectrum, in an experiment similar to that of Ref.~\cite{greiner}.

Summarizing, we have proposed an experimentally feasible and relatively simple
method of creating a system whose physics is governed by the
disordered Bose Hubbard model. We have shown how  the onset of
small perturbation
of the lattice potential may result in a
dynamical transition from a superfluid regime into Bose-glass, or Anderson-localized phases.
This transition occurs within experimentally feasible time scales
and can be easily controlled, allowing for
a detailed analysis of disorder induced transitions.
Our proposal stimulates in this sense new interesting experimental
possibilities including studies of Anderson localization in 2D systems
(which is still a controversial topic),
and the investigation of the SF to MI transition in the presence of disorder.

We acknowledge support from the Alexander von Humboldt Stiftung, the
Deutscher Akademischer Austauschdienst (DAAD), the Deutsche
Forschungsgemeinschaft, EU RTN Network "Cold Quantum Gases",
and ESF PESC Program BEC2000+. Work of B.D. and J.Z. was partially supported
by KBN under projects 2 P03B 124 22 and 5 P03B 088 21, respectively.
We thank Krzysztof Sacha and Misha Baranov for discussions.


\begin{references}
\bibitem{remark}
The same quantitative results are obtained in  the model with a
truely randomly  generated disorder.
\bibitem{notka1} Exactly the same conclusion can be formulated in a 2D case depicted in
Fig.~\ref{spec}.
\end{references}
\end{document}